\documentclass{PoS}

\title{MeerTime - the MeerKAT Key Science Program on Pulsar Timing}

\ShortTitle{MeerTime}

\author{
M. Bailes$^{\dagger a}$,
E. Barr$^b$, %Not provided
N. D. R. Bhat$^c$,
J. Brink$^d$,
S. Buchner$^e$,
M. Burgay$^f$,
F. Camilo$^e$,
D. J. Champion$^b$,
J. Hessels$^g$,
G. H. Janssen$^{g,h}$,
A. Jameson$^a$,
S. Johnston$^i$,
A. Karastergiou$^j$,
R. Karuppusamy$^b$, %Not provided
V. Kaspi$^k$,
M. J. Keith$^l$,
M. Kramer$^b$,
M. A. McLaughlin$^m$,
K. Moodley$^n$,
S. Oslowski$^a$,
A. Possenti$^f$,
S. M. Ransom$^o$,
F. A. Rasio$^p$,
J. Sievers$^q$,
M. Serylak$^e$,
B. W. Stappers$^l$,
I. H. Stairs$^r$,
G. Theureau$^{stu}$,
W. van Straten$^v$,
P. Weltevrede$^l$,
N. Wex$^b$\\ %Not provided
\llap{$^\dagger$} - Speaker.\\
\llap{$^a$}OzGrav, Centre for Astrophysics and Supercomputing, 
Swinburne University of Technology, H11 PO Box 218 Hawthorn, Vic, 3122, Australia\\
\llap{$^b$}Max-Planck-Institut f\"ur Radioastronomie  Auf dem H\"ugel 69, D-53121 Bonn, Germany\\
\llap{$^c$}International Centre for Radio Astronomy Research, Curtin University Bentley, WA 6102, Australia\\
\llap{$^d$}Independent African P.O. Box 17633, Bainsvlei, South Africa, 9338\\
\llap{$^e$}Square Kilometer Array South Africa, The Park, Park Road, Pinelands, Cape Town 7405\\
\llap{$^f$}INAF - Osservatorio Astronomico di Cagliari via della Scienza 5, 09047 Selargius (CA), Italy\\
\llap{$^g$} ASTRON, the Netherlands Institute for Radio Astronomy, Postbus 2, 7990 AA, Dwingeloo, The Netherlands.\\
\llap{$^h$} Department of Astrophysics/IMAPP, Radboud University, P.O. Box 9010, 6500 GL Nijmegen, The Netherlands.\\
\llap{$^i$} CSIRO Astronomy and Space Science PO BOX 76, NSW 1710, Australia.\\
\llap{$^j$} Astrophysics, University of Oxford, Denys Wilkinson Building, Keble Road, Oxford OX1 3RH, UK\\
\llap{$^k$} McGill University, 3600 University St., Montreal, QC H3A 2T8, Canada\\
\llap{$^l$}Jodrell Bank Centre for Astrophysics, School of Physics , Astronomy, University of Manchester Alan Turing Building, Oxford Road, Manchester, M13 9PL, UK\\
\llap{$^m$} WVU, Center for Gravitational Waves and Cosmology, Chestnut Ridge Research Building, Morgantown, WV 26506\\
\llap{$^n$} Astrophysics and Cosmology Research Unit, School of Mathematics, Statistics and Computer Science, University of KwaZulu-Natal, Durban, 4041, South Africa\\
\llap{$^o$}NRAO,  520 Edgemont Rd., Charlottesville, VA, 22903, USA\\
\llap{$^q$}Astrophysics , Cosmology Research Unit, School of Mathematics, Statistics , Computer Science, University of KwaZulu-Natal, Durban, 4041, South Africa\\
\llap{$^r$} Dept. of Physics and Astronomy, University of British Columbia, 6224 Agricultural Road, Vancouver, BC V6T 1Z1 Canada\\
\llap{$^p$}CIERA, Northwestern University, Evanston, IL, USA\\
\llap{$^r$}Dept. of Physics , Astronomy, University of British Columbia, 6224 Agricultural Road, Vancouver, BC V6T 1Z1 Canada\\
\llap{$^s$}Laboratoire de Physique et Chimie de l'Environnement et de l'Espace LPC2E CNRS-Universit{\'e} d'Orl{\'e}ans, F-45071 Orl{\'e}ans, France\\
\llap{$^t$}Station de radioastronomie de Nan{\c c}ay, Observatoire de Paris, PSL Research University, CNRS/INSU F-18330 Nan{\c c}ay, France\\
\llap{$^u$}Laboratoire Univers et Th\'eories LUTh, Observatoire de Paris, PSL Research University, CNRS/INSU, Universit{\'e} Paris Diderot, 5 place Jules Janssen, 92190 Meudon, France\\
\llap{$^v$}Institute for Radio Astronomy \& Space Research, Auckland University of Technology, Private Bag 92006, Auckl, 1142, New Zealand,\\
}

\abstract{
The MeerKAT telescope represents an outstanding opportunity for radio pulsar timing 
science with its unique combination of a large collecting area and aperture efficiency (effective
area $\sim$7500 m$^2$), system temperature ($T<20$K), 
high slew speeds (1-2 deg/s), large bandwidths (770 MHz at 20cm wavelengths), 
southern hemisphere location (latitude $\sim -30^\circ$) and 
ability to form up to four sub-arrays. The MeerTime project is a five-year program on the MeerKAT array by
an international consortium that will regularly time
over 1000 radio pulsars to perform tests of relativistic gravity, search for the gravitational wave
signature induced by supermassive black hole binaries in the timing residuals of millisecond
pulsars, explore the interiors of neutron stars through a pulsar glitch monitoring programme, 
explore the origin and evolution of binary pulsars, monitor the swarms of pulsars that inhabit
globular clusters and monitor radio magnetars. MeerTime will complement the TRAPUM project
and time pulsars TRAPUM discovers in surveys of the galactic plane, globular clusters and the galactic
centre. In addition to these primary programmes, over 1000 pulsars will have their arrival
times monitored and the data made immediately public.
The MeerTime pulsar backend comprises two server-class machines each of which possess 
four Graphics Processing Units. Up to four pulsars can be coherently dedispersed simultaneously up
to dispersion measures of over 1000 pc cm$^{-3}$. All data will be provided in psrfits format.
The MeerTime backend will be capable of producing coherently dedispersed filterbank
data for timing multiple pulsars in the cores of globular clusters that is useful for pulsar searches
of tied array beams.
The first real-time pulsar profiles have been obtained as part of the MeerKAT commissioning
process, and useful scientific data will start to come online through 2017. All MeerTime data
will ultimately be made available for public use, and any published results will include
the arrival times and profiles used in the results.
}

\FullConference{MeerKAT Science: On the Pathway to the SKA\\
                 25-27 May, 2016\\
                 Stellenbosch, South Africa}

\begin{document}

\section{Introduction}

Fifty years after their discovery[1]\nocite{1968Natur.217..709H}, radio pulsars remain a fertile area of astronomical research. In the standard model[2]\nocite{2004hpa..book.....L}, radio pulsars are highly-magnetised neutron stars with magnetic field strengths of 10$^8$-$10^{14}$G, some 10 km in radius and with masses of between $\sim$ 1.2-2 M$_\odot$. Magnetic dipole radiation removes energy from the pulsar, and a light-house beam of radio emission sweeps
through space and results in a train of regular pulses that can be routinely detected by
radio telescopes of sufficient aperture. The energy loss results in a slow spin-down
of the pulsar, and by accurately measuring pulse times of arrival at the observatory, a
large number of observable parameters become available to pulsar astronomers.

The MeerKAT Key Science Project on Pulsar Timing originated in 2010 when a strong science case was put to
an international time allocation committee based upon the then draft sensitivity for the MeerKAT telescope of
220 m$^2$/K and bandwidth of 850 MHz. The project received a draft allocation of 7800h and the 
Key Science Project is now referred to as ``MeerTime''.

\subsection{Background}
Up until 1974 all known pulsars were solitary objects with spin periods $P>33$ ms with limited astrophysical applications. 
Two major discoveries rocked the pulsar community when improved pulsar survey instrumentation enabled the discovery of faster pulsars. In 1974 the binary PSR B1913+16 was found using the Arecibo 
telescope[3]\nocite{1975ApJ...195L..51H}. It was a 59 ms pulsar orbiting another neutron star every 7.75 hours. Its timing ultimately led to the Nobel prize in physics for verifying the existence of gravitational waves. Then in 1981 the celebrated 1.55 millisecond pulsar, PSR B1937+21 was discovered [4] \nocite{1982Natur.300..615B}. Today the pulsar catalogue[5] \nocite{2005AJ....129.1993M} lists over 250 binary pulsars (almost 10\% of the population) and 350 millisecond pulsars. These pulsars are ideal relativistic laboratories and probes of stellar evolution. Their inherent stability give them applications in the search for gravitational waves. The majority of known pulsars reside in the southern hemisphere and 95\% of all pulsars are accessible to
MeerKAT with its Southern location and 75 degree elevation limit.

\subsection{MeerTime Science Themes}
MeerTime has three high-priority science themes:

\begin{itemize}
\item
{\bf Relativistic and Binary Pulsars: } Pulsars that possess compact binary companions in tight relativistic orbits allow tests of General Relativity and its alternatives that are impossible to perform elsewhere. Binary pulsars are also fossil records of stellar evolution and accretion physics. In addition General Relativity can be used to determine the masses of millisecond pulsars that inform us about the equation of state of nuclear matter.

\item
{\bf Millisecond Pulsar Timing and Gravitational Wave Detection:} Millisecond pulsars can be used to search for the signatures of gravitational waves generated by supermassive black hole binaries and/or cosmic strings in the early Universe. 

\item
{\bf Globular Cluster Pulsar Timing:} Globular clusters are breeding grounds for millisecond pulsars and via partner exchange in their dense cores produce exotic systems that permit experiments otherwise impossible to perform that inform us about General Relativity and also the equation of state of nuclear matter via neutron star spins and masses. 
\end{itemize}

MeerKAT's design (comprising of 64 independent elements) means that it is possible to form up to four sub-arrays, 
allowing for some novel approaches to pulsar timing. This permits sub-array modes that enable the simultaneous
production of pulse arrival times from four different pulsars in different regions of the sky.

The MeerKAT system design originally forecast a sensitivity of some 220 m$^2$/K, but it
now appears that the aperture efficiency and system temperature may enable a remarkable sensitivity
in excess of 400 m$^2$/K, virtually quadrupling its efficiency for pulsar timing. For reference the 
other Southern hemisphere radio telescopes are all less than 70 m$^2$/K, although the
new ultra-wideband receiver (700 MHz-4 GHz) for the Parkes 64\,m radio telescope still make it a 
highly-effective pulsar telescope, particularly for dispersion measure determinations.

Further classes of pulsar were part of the original science case:

\begin{itemize}
\item
{\bf Young, Glitching and Highly-magnetised pulsars:} These objects allow probes of neutron star interiors where physics is at its most extreme. 
\item
{\bf The Thousand Pulsar Array:} Most pulsars are not timed regularly because of the time commitment required on otherwise over-subscribed instruments. It can be demonstrated that MeerKAT can time over 1000 pulsars a day, and that for a modest increase in MeerTime's time request a legacy dataset could be created that monitored most of the known pulsar population. This would answer questions like, do ``normal'' pulsars also glitch? 
Are the period derivatives stable? What are the proper motions of pulsars? 
Are pulsar pulse profiles really stable? 
If we monitor a thousand years of pulsar rotation history every year, what unexpected phenomena will we see?
\item
{\bf RRATs (Rotating RAdio Transients):} In 2010 the newly discovered population of ``RRATs'' was a topic of much interest. These objects appeared to be pulsars that only pulsed sporadically. Their birthrate appeared extremely high, yet little was known about the population. This population was the final science theme of the original proposal.
\end{itemize}

The construction of the MeerKAT telescope is a great opportunity to advance radio pulsar
science. The vast majority of pulsars ($>95$\%) are visible from the MeerKAT
and its combination of a large collecting area, cool receivers, rapid slew speed
and wide bandwidth makes it a powerful facility for the great plethora of pulsar science.

In this paper we describe the MeerTime pulsar timing experiment. In section 2
we take the telescope's technical specifications and determine the speed with
which it can time the radio pulsar population before discussing the pulsar
backend subsystem in section 3. Finally, in section 4, we discuss the main science cases
for the telescope.

\section{The MeerKAT as a pulsar telescope}

Although individual pulses from a pulsar are irregular in their amplitude, shape
and polarisation, their mean profile is often remarkably stable. This fact
permits extremely accurate pulse arrival times (ToAs) to be derived by integrating
a pulsar's pulses for many 100s-thousands of periods and comparing the time-tagged
profile against a standard. The error in a pulse ToA $\sigma$ depends upon the time derivative
of the profile and the signal-to-noise ratio $snr$. A good rule of thumb is that the error
in a ToA $\sigma$ can be approximated by
\begin{equation}
\sigma \sim w/(2 \times snr) 
\end{equation} 
where $w$ is the half width of the profile. In reality, this relation is only an approximation
and assumes that every pulse is identical. A more correct approach would use the derivative
of the pulse profile, and make some allowance for the fact that not every pulse is identical.

The radiometer equation for a radio telescope describes the $snr$\, expected from
a source of flux density $S$ as
\begin{equation}
snr = {{S G \sqrt{(B N_{\rm p} t)}}\over{T_{\rm rec}+T_{\rm sky}}}\sqrt{{{P-w}\over{w}}}
\end{equation}
\noindent 
where $G$ is the gain of the telescope in K Jy$^{-1}$, $B$ is the processed bandwidth (in Hz), $N_{\rm p}$ is
the number of orthogonal polarisations (maximum of 2), $t$ is the integration time in seconds, $T_{\rm rec}$
and $T_{\rm sky}$ are the receiver and system temperature respectively (in K), $P$ is the pulse
period and $w$ is the width of the pulse.

The first determination of the MeerKAT's technical specifications are extremely encouraging.
The dishes are an offset gregorian and have an effective diameter of 13.965\,m and an aperture
efficiency of 0.71-0.81. If we adopt 0.76 as a mean efficiency this gives each dish a gain $G_0$ of
\begin{equation}
G_0=0.76 \pi {{(13.965/2)^2}\over{2k}} = 0.042 {\rm \,K\, Jy}^{-1}
\end{equation}
where $k$ is Boltzmann's constant. If added coherently, the 64-dish instrument has a total gain of
$G=64 G_0 = 2.7$ K Jy$^{-1}$.

In array release 3, the telescope will possess a total usable bandwidth of 770 MHz of bandwidth in two
orthogonal polarisations. Early tests of the receivers suggests that $T_{\rm rec}$ will
be less than 20K, possibly as low as $\sim$15K. 

If MeerKAT observes the 2 mJy millisecond pulsar PSR J1909-3744, which has a pulse width of
40 $\mu$s for just 15m, the expected $snr \sim $2700 (pessimistically assuming
$T_{\rm rec}=20$K) and the ToA error just $\sim$8 ns (in reality there is likely to be pulse jitter
that limits such accuracies being obtained).

Rather than slewing all over the sky looking for MSPs that are currently experiencing scintillation
maxima with cumbersome 64-100m-class telescopes, the MeerKAT will be able to split off a sub-array
and quickly determine which MSPs are in a ``bright state'' to further improve observing efficiency with
its impressive slew speeds of 1-2 deg/s.

\section{The MeerKAT pulsar backend system}

In the mid-late 1990s pulsar astronomers recognised the increases in timing precision that could be achieved by coherently dedispersing the voltages induced in the receiver of a single dish by the deconvolution of the signal with an appropriate filter. Coherent dedispersion used to represent a major challenge to pulsar astronomy, requiring custom boards to digitise the voltages at the requisite rate (1/$B$) where $B$ is the bandwidth of the backend and capture them in clusters of computers. It was originally thought that a computing
cluster might require a large cluster of workstations to achieve the necessary signal processing
to coherently dedisperse the radio frequency signals from MeerKAT, but since 2010 there have been several important developments in the creation of coherent dedispersion pulsar processors or ``backends''. 

\begin{itemize}
\item
The community has developed open-source software that has been largely adopted providing increased rigour and testing. Most of the pulsar community now uses and contributes to the psrchive suite of tools to process and manipulate folded pulsar profile data [6] \nocite{2004PASA...21..302H}. %(Hotan, van Straten \& Manchester 2004, PASA 21, 304). 
\item
A software library (psrdada) that both captures data from UDP streams and moves it to computer's memory (RAM) is available from sourceforge.com that permits the quick development of capture engines. A library (dspsr) that transforms voltage data to coherently-dedispersed folded profiles is also available as an open-source software library [7] \nocite{2011PASA...28....1V} %(van Straten \& Bailes 2010, PASA 28, 1). 
These open-source libraries greatly facilitate the creation of pulsar processing backends.
\item
Large-N Fourier transforms used to be extremely expensive to compute, requiring vast arrays of CPUs and multiplexing of the data because of the time taken to compute each one. The technological breakthrough of the graphics processing unit (GPU) in consumer games cards has reduced the cost of the necessary computations by more than an order of magnitude since 2010. Almost all pulsar processors now utilise GPUs as the Fourier transform engine.
\item
Back in 2010, 10-Gb ethernet represented the peak performance one could aspire to when trying to perform lossless data capture. This meant that with 8-bit sampling, several O(4) computers would be required to capture a single stream from MeerKAT's 
tied array beams. Now it is possible to have a single machine capture over 54 Gb/s of data without loss using dual
40 Gb Network Interface Cards (NICs).
\item
The use of interferometers for pulsar timing has been steadily increasing. The Westerbork array has been joined by the LEAP project that ties the major European VLBI telescopes into a single coherent beam, and the Very Large Array, LOFAR and the UTMOST project in Australia are all examples of interferometers observing pulsars.
\end{itemize}

MeerKAT's pulsar timing hardware comprises two machines that were used to prototype the pulsar processor for the SKA at Swinburne University of Technology. These machines can coherently dedisperse two parallel 850 MHz dual-polarisation streams simultaneously and one is currently at the MeerKAT site. 

Progress towards regular pulsar timing is continuing. A major breakthrough occurred in Q2 2016 when the first pulsar profile from the Vela pulsar (PSR J0835-4510) was produced from a single beam with data written to disk. 
This observation confirmed that MeerKAT single dishes were producing very high quality data with system temperatures near the published specification (<20K).  Shortly afterwards, the first tied array beam was created on the bright millisecond pulsar PSR J0437--4715 and processed in software from voltages recorded to disk.  Although satellite transmissions are present in the band, 
over 75\% of it is ``useable''. In October the first real-time pulsar profiles were produced validating that the pulsar processor can capture data at the requisite rate and process them. Currently the polyphase filterbank data cannot be correctly coherently dedispersed because of digitally-induced artefacts in the polyphase filterbank frequency channels. Coherent dedispersion will be required to achieve the ultimate timing precision promised by the system and should be possible when the beam-former and polyphase filterbanks are moved to the SKARAB boards in mid 2017.

\subsection{The Max Planck Institute 1.7-3.5 GHz receiver upgrade}

High dispersion pulsars suffer from multi-path propagation effects that limit their use in pulsar timing experiments. The effects scale as the wavelength to the power 4.4, and hence some pulsars demand the use of the highest frequencies to perform the best science. Much beyond 3 GHz there are relatively few pulsars that still retain enough flux to make this worthwhile, as many pulsars have steep spectral indices (usually between --1 and --3).

Professor Michael Kramer (MPIfR) has been leading a project to upgrade MeerKAT to operate beyond 1.7 GHz with the deployment of 64$\times$1.7-3.5 GHz receivers. This project will enable both wider bandwidths and reduced scattering on many of our higher dispersion measure pulsars. It was \nocite{2015Sci...349.1522S} 
recently demonstrated [8] that some pulsars (e.g. PSR J1909--3744) produce arrival times of exceptional accuracy at these frequencies that are critical for the direct detection of gravitational waves. 

By sub-arraying, it will be possible for us to observe pulsars all the way from 0.9 to 3.5 GHz, albeit
with reduced gain because the telescope will have to be ``sub-arrayed''. This will provide an exceptional lever arm to define the variable dispersion measures (a pulsar timing pollutant). When the UHF receivers are in place this can extend all the way from 0.5-3.5 GHz.

\section{The MeerTime Science Case} 
\subsection{Millisecond Pulsar Timing}

In MeerTime's original 2010 science case it was anticipated that MeerKAT could independently detect a gravitational wave background after 5 years if the dimensional amplitude exceeded $2\times10^{-15}$
based upon limits in vogue at the time [9]. \nocite{2009MNRAS.400..951V} 
The current best millisecond pulsar for timing accuracy (PSR J1909--3744) already suggests that an amplitude of this magnitude is ruled out[8] \nocite{2015Sci...349.1522S} and that gravitational wave detection from pulsars will require international coordination and cooperation. MeerKAT can dramatically increase the pool of MSPs from which a gravitational wave background or individual binaries can be searched for with its unique combination of sensitivity, geographical location, ability to sub-array and the speed at which it can traverse the sky. Reardon et al. (2016) [10]
\nocite{2016MNRAS.455.1751R} recently reported on the timing of 20 MSPs from Parkes and
based upon his residuals and the relative sensitivity of the two telescopes, the MeerKAT should increase the
number of pulsars with sub-us residuals from 5 to 16 objects if sensitivity was the only improvement factor,
but MeerKAT's ability to subarray and seek out those MSPs that are experiencing scintillation maxima
gives us hope that it can do much better than a simple scaling of sensitivities might suggest.

Several studies [11,12,13,14] \nocite{2011MNRAS.417.2916L, 2011MNRAS.418.1258O, 2013MNRAS.430..416O, 2014MNRAS.443.1463S}  
have demonstrated that pulsar timing precision is ultimately limited by the stochastic wideband 
impulse-modulated self-noise (SWIMS, also known as jitter and single-pulse variability) that is intrinsic 
to the pulsar emission.  Consequently, optimal use of full array sensitivity requires the ability to divide 
it into sub-arrays; furthermore, because it is imperative to account for this noise in high-precision pulsar timing data analysis, the instrumentation for pulsar timing must be updated to produce additional statistical information.
\nocite{2011MNRAS.418.1258O, 2013MNRAS.430..416O} 
 It has been demonstrated[12,13] that arrival time estimation bias can be mitigated by measuring the periodic correlations of the Stokes parameters and Shannon et al. (2014) [14] \nocite{2014MNRAS.443.1463S} have described how jitter noise can be characterised and incorporated in estimates of arrival time precision.  Ongoing research by our team will combine these approaches using generalised least squares estimation to simultaneously reduce bias, accurately estimate uncertainty, and increase the sensitivity of experiments such as pulsar timing arrays.

Gravitational waves are just one of the exciting science cases to be realised by timing an array of millisecond pulsars. Timing residuals also contain a wealth of information about the parameters of the parent binary, useful for studies of stellar evolution, the IGM and even our own planetary ephemerides. Since our original proposal the Fermi satellite has unveiled a tremendous population of millisecond pulsars (now up to 350) waiting for an instrument capable of producing accurate arrival times to capitalise on them. MeerKAT is such an instrument.

\subsection{Relativistic and Binary Pulsar Timing}
						
Pulsars are remarkable laboratories for the study of gravitation. In both the highly relativistic interior and the vicinity of a pulsar (and its binary companion, in case of double neutron-star systems or potential pulsar-black hole system) space-time may significantly deviate from the predictions of General Relativity (GR) [15].\nocite{1996PhRvD..54.1474D}
%(Damour \& Esposito-Farèse 1996, PhRvD 54, 1474). 
Pulsar timing therefore provides a unique tool for probing gravity in the strong field regime, 
enabling high-precision tests of GR or other theories of gravity. Double-neutron-star systems such as the Double Pulsar [16]
\nocite{2006Sci...314...97K}
%(Kramer et al. 2006, Sci 314, 97) 
provide unrivalled probes for testing most aspects of GR.  
Binary pulsars with a white dwarf companion and hence large mass dipole can set interesting constraints on 
alternative theories that predict, for instance, the existence of gravitational dipole radiation [17].\nocite{2013Sci...340..448A}
%(e.g. Antoniadis et al. 2013, Sci 340, 448).  
Resolving tight binary orbits to investigate effects such as 
the Shapiro delay requires short-spaced observations with high sensitivity.  Meanwhile, identifying the 
weak signatures of subtle relativistic effects needs long-term monitoring with good cadence.   MeerKAT's 
excellent sensitivity (surpassing even our high expectations in 2010) and good frequency coverage will make it the premier telescope for studying Southern-sky pulsars. This will not only improve existing GR tests but will allow us to measure new effects to probe new physics. This is best demonstrated with the unique Double Pulsar, where the precision of tests of gravity will go beyond the current best weak-field tests in the solar system. As has been shown
(Kehl 2015, Masters thesis, University of Bonn), 
we expect to measure the moment-of-inertia of the J0737-3039A in the Double Pulsar 
for the first time, providing a handle on the equation-of-state of super-dense matter. 

With the sensitivity provided by MeerTime, we will also determine masses for both pulsars and their companions. These can be used to test theories of binary evolution\nocite{1999A&A...350..928T} [18]
% (e.g. Tauris \& Savonije 1999, A\&A, 350, 928) 
and 
to investigate the distribution of neutron-star masses.  In particular, the discovery of massive neutron stars [19,17]
\nocite{2010Natur.467.1081D, 2013Sci...340..448A}
%(Demorest et al. 2010, Nature 467, 1081; Antoniadis et al. 2013) 
suggests that high-mass population of neutron stars exists, even possibly resulting from birth [20,21].
\nocite{2011MNRAS.416.2130T, 2016ApJ...830...36A}
%(e.g. Tauris et al. 2011 MNRAS, 416, 2130, Antoniadis et al 2016, submitted, arXiv:1605.01665).  
As mass statistics improve, we will get closer to identifying the maximum mass possible for a neutron star, itself a constraint on the equation of state.  

MeerTime will furthermore provide astrometry (distances, proper motions and hence velocities) for millisecond and binary pulsars, allowing us to infer their birth velocities and constrain asymmetric supernova kicks, particularly in double-neutron-star systems[22].\nocite{2006MNRAS.373L..50S}
%(e.g., Stairs et al 2006, MNRAS 373, L50).

\subsection{ Globular Cluster Pulsar Timing}

Globular clusters are treasure troves of exotic millisecond pulsars,
for a recent scientific overview see reference [23].
\nocite{2015aska.confE..47H}
The cores of globular clusters have stellar densities 10$^3$ - 10$^4$ times greater than in the Galactic field; this promotes the formation of binary systems in which a neutron star can be recycled to millisecond rotation rates via the transfer of matter and angular momentum from a Roche-lobe-filling companion.  This extreme stellar density can also lead to exchange interactions, which create bizarre pulsar systems, unlike anything so-far seen in the Galactic field.  Currently there are 
146 pulsars known in 28 globular clusters - including the fastest-spinning pulsar known[24], \nocite{2006Sci...311.1901H}
%(Hessels et al. 2006, Sci 311, 1901), 
exotic eccentric binaries suitable for neutron star mass measurements [25] \nocite{2005Sci...307..892R}
%(Ransom et al. 2005, Sci 307, 892), 
and a unique triple system with a planetary companion [26].\nocite{2003Sci...301..193S}. 
%(Sigurdsson et al. 2003, Sci 301, 193).

Literally all of these 146 pulsars are visible to MeerKAT and MeerTime plans a sensitive, and comprehensive globular cluster pulsar timing campaign.  Combining MeerKAT timing data with up to three decades of archival measurements from GBT, Arecibo, and Parkes, MeerTime will probe the spin, orbital, and proper motions of these pulsars in unprecedented detail and measure previously inaccessible system parameters that will allow us to probe accretion physics, dense matter, gravitational theories, and the evolution and properties of the clusters themselves in exquisite detail.  From a practical point of view, timing globular cluster pulsars also provides a great efficency because in some cases (e.g. M28, 47 Tucanae and Terzan 5) dozens of millisecond pulsars can be observed simultaneously.  MeerKAT will revolutionize searches of southern globular clusters via TRAPUM and the long-term timing of these and existing pulsars via MeerTime.  To achieve sensitivity to ~10 $\mu$Jy pulsars in these 
clusters, MeerTime plans typically 1-hr timing sessions for these clusters. 

\subsection{The Thousand Pulsar Array}

Not all pulsars pulse regularly. Since the connection between pulsar radio emission and timing properties in
the so-called intermittent pulsars was first pointed out [27] \nocite{2006Sci...312..549K} more pulsars
exhibiting these properties have been discovered, as well as, e.g., multi-wavelength moding pulsars where the pulse profile changes significantly between two states having different radio and X-ray properties [28].
\nocite{2013Sci...339..436H}
Secular variations in previously thought stable pulse profiles were also seen for a large sample of ordinary pulsars [29] \nocite{2010Sci...329..408L} with a clear connection between timing irregularities/noise and emission properties/profiles. In this context, progress is being made on understanding pulsar interiors and how/if the observed pulse profile variations could be ascribed to long term free precession. Finally, the LOFAR telescope is revealing the complex imprints of the ISM on pulsar data [30]. \nocite{2014ApJ...790L..22A}
The Thousand Pulsar Timing Array will provide an opportunity to study the breadth of pulsar phenomenology. That will result in new breakthroughs relating to the interiors, to the magnetosphere and to the environment of pulsars, to ISM and Galactic magnetic field studies, and will lead to improved pulsar timing. MeerKAT is an exceptionally sensitive telescope for this purpose. We estimate that, with the current full MeerKAT sensitivity, we can obtain a high signal to noise (>20) profile of a pulsar with a duty cycle of 5\% and a flux density of 0.30 mJy (of which more than 1000 are visible from Meerkat) within a minute of observation. From a sensitivity perspective therefore, it is easy to accommodate regular observations of 1000 pulsars within the requested 16 h per observing epoch. 

\subsection{The Timing of Young and Energetic Pulsars}

Young and energetic pulsars are often associated with supernova remnants (SNRs), pulsar wind nebulae (PWNe), and/or high-energy X-ray/gamma-ray point sources [31]. \nocite{2002ApJ...577L..19R}
%(e.g. Roberts et al. 2002, ApJ 577, 19).  
Timing of young pulsars provides their spin-down rate, which then sets the energy budget powering the PWN and other high-energy emission [32]. \nocite{2002ApJ...579L..25C}
%(e.g., Camilo et al. 2002, ApJ, 579, 25).  
Long-term timing provides the proper motion, which is a key ingredient for deciphering the morphology of SNRs and PWNe (including bow shocks).  In some cases, it is also possible to measure the neutron star's braking index, revealing the multi-pole nature of the magnetic field and perhaps also its evolution in time [33]. \nocite{2005ApJ...619.1046L}
%(e.g., Livingstone et al. 2005, ApJ, 619, 1046).  
Many young pulsars also show glitches, which probe the neutron star interior in a unique way [34].
\nocite{2011MNRAS.414.1679E}
%(Espinoza et al. 2011, MNRAS, 414, 1679).  
With MeerKAT's large sensitivity, it may also be possible to probe the nebulae surrounding some young pulsars through precise characterization of their scattering/dispersion/rotation measure with time or other propagation effects [35]. \nocite{2001MNRAS.321...67L}
%(e.g. Lyne et al. 2001, MNRAS, 321, 67).

\subsection{ Magnetar Pulsar Timing}

To date, 4 of 23 known magnetars have been detected at radio wavelengths [36].\nocite{2006Natur.442..892C}
%(e.g., Camilo et al. 2006, Nature, 442, 892). 
They share some characteristics distinct from those of the normal pulsar population, such as flat radio spectra. They are also extremely variable; 2 of the 4 are currently no longer radio emitters [37] \nocite{2016ApJ...820..110C}
%(e.g., Camilo et al. 2016, ApJ, 820, 110), 
although this could change, and new ones could be discovered. Through frequent timing observations of the active radio magnetars we aim to obtain a continuous record of their torque, which illuminates the continued release of magnetic energy in the neutron star. Our broader aims are to develop a better understanding of the dynamical behaviour of magnetar magnetospheres, and to establish the conditions under which radio emission takes place therein. Very frequent observations are needed because the torque on a magnetar can change by ~10\% on ~weekly timescales.

\subsection{The Rotating Radio Transients}

These objects are no longer being observed as part of our timing project. Their poor positions
are not well-mapped to the small tied array beam of MeerTime and they don't appear to
be anything except an extension of nulling pulsars.

\section{Observing strategy and data products}

To detect the stochastic gravitational wave background or individual sources requires the highest precision possible. This calls for regular observing cadence, preferably of order 20 times per year.
These observations will make MeerKAT a critical contributor to the broader International Pulsar Timing Array (IPTA) effort.
MeerKAT has a unique opportunity to contribute to the direct detection of gravitational waves. 
It will be the most sensitive telescope in the Southern hemisphere, and as mentioned before its ability to sub-array means it can employ novel techniques to fully exploit MSP scintillation. It will be possible for sub-arrays to be searching for MSPs to time that are at scintillation maxima, whilst the majority of the antennas are conducting routine timing.  

The best observing strategy for binary and millisecond pulsars is to observe entire orbits where practical (this prevents unfortunate covariances between binary and other parameters) and to have occasional ``campaigns'' when the cusps in Shapiro delays are visible. MeerTime's immediate focus is on systems in which gravitational wave emission is implied by monitoring their orbital period derivatives and on those binaries where pulsar masses can be achieved with MeerKAT's increased sensitivity over existing facilities.
Currently, orbital decays have been detected for 8 binary pulsar systems 
while detections for a further 4-6 (some unpublished pulsars) systems
can be expected with MeerKAT. In addition, mass measurements of binary neutron 
stars should be possible for at least 10  more systems [38].\nocite{2016ARA&A..54..401O}
Full orbits on these systems are possible with MeerKAT but not telescopes like Arecibo or FAST. 

Most pulsars in globular clusters are very stable timers. As one of the aimed measurables is the proper motion of clusters, many sessions throughout the year are necessary. Since globular cluster MSPs are in average fainter sources than the Galactic ones, long integrations are a necessity.

To enable science it is essential that the pulse profiles produced by the backend are created in a format accessible via public domain packages such as psrchive. The pulsar processor will create FITS-format folded archives and coherent filterbanks at a nominal dispersion measure, also in FITS. The arrival times will be in reference to the observatory clock, which will ultimately be referenced to UTC-NIST. 

The dimension of the scientific data products is manageable (60 TB over five years). Software pipelines will be made open access and available on the project website, as will clock correction files. 

Once the data are calibrated they can be fit with the standard pulsar timing packages tempo, 
tempo2 and PINT.

It is rare that a single epoch of pulsar timing results in a publishable outcome. Instead pulsar parameters slowly become scientifically interesting as the time span increases. Pulsar positions rapidly increase their precision once a year of data is obtained, proper motions usually take a few years to manifest themselves in timing residuals and the discovery of the gravitational wave background is likely to take a decade or so. MeerTime's data release policy is as follows: Data from the 1000 pulsar array will be available immediately. The MSP, relativistic binary and globular pulsar data will be released 18 months after they are recorded. We intend to publish all times of arrival on something like a annual basis in scheduled ``data releases''. All publications will provide the arrival times and raw data from the observations that led to any claimed results.

\section*{Acknowledgements}
M Bailes acknowledges the Australian Research Council grants 
CE170100004 (OzGrav) and FL150100148.
A. P. and M. Bu. acknowledge the support of the Italian Ministry of Foreign Affairs and International Cooperation, Directorate General for the Country Promotion (Bilateral Grant Agreement ZA14GR02 - Mapping the Universe on the Pathway to SKA). This research is supported by the Max-Planck-Society and by the ERC Synergy Grant ``BlackHoleCam: Imaging the Event Horizon of Black Holes'' (Grant 610058). Pulsar research at UBC is supported by an NSERC Discovery Grant and by the Canadian Institute for Advanced Research. This research is supported by NSF IRES Award 7706217.
\bibliographystyle{ws-procs975x65}
\bibliography{journals,meertime}
\end{document}